\documentclass[12pt]{iopart}

\usepackage{graphicx}
\usepackage{dcolumn}
\usepackage{bm}
\usepackage{multirow}
\usepackage{iopams}
\usepackage{soul}
\usepackage{epsfig} 
\usepackage{lscape}
\usepackage{color}

\begin{document}


\title{Spreading of diseases through comorbidity networks across life and gender}

\author{Anna  Chmiel$^1$, Peter Klimek$^1$, Stefan Thurner$^{1,2,3}$}
\ead{stefan.thurner@meduniwien.ac.at}
\address{
$^1$Section for Science of Complex Systems; CeMSIIS; Medical University of Vienna; Spitalgasse 23; A-1090; Austria. 
specified
$^2$Santa Fe Institute; 1399 Hyde Park Road; Santa Fe; NM 87501; USA. 
$^3$IIASA, Schlossplatz 1, A-2361 Laxenburg; Austria.
}%
\begin{abstract}
The state of health of patients is typically not characterized  by a single disease alone but by multiple (comorbid) medical conditions. 
These comorbidities may  depend strongly on age and gender. We propose a specific phenomenological comorbidity network of 
human diseases that is based on medical claims data of the entire population of Austria. The network is constructed from a two-layer 
multiplex network, where in one layer the links represent the conditional probability for a comorbidity, and in the other the links contain 
the respective statistical significance. We show that the network undergoes dramatic structural changes across the lifetime of patients.
Disease networks for children consist of a single, strongly inter-connected cluster. During adolescence and adulthood further disease 
clusters emerge that are related to specific classes of diseases, such as circulatory, mental, or genitourinary disorders.
For people above 65 these clusters start to merge and highly connected hubs dominate the network. These hubs are related to 
hypertension, chronic ischemic heart diseases, and chronic obstructive pulmonary diseases. We introduce a simple diffusion model to 
understand the spreading of diseases on the disease network at the population level. For the first time we are able to show that patients 
predominantly develop diseases which are in close network-proximity to disorders that they already suffer. The model explains more 
than 85\% of the variance of all disease incidents in the population. 
The presented methodology could be of importance for anticipating age-dependent disease-profiles for entire populations, and for 
validation and of prevention schemes. 
\end{abstract}
\pacs{{\color{blue} 89.75.Fb, 87.19.X-, 87.85.Tu}}

\maketitle

\section{\label{sec:intro}Introduction}

Diseases are usually defined by a set of phenotypes that are associated with various pathobiological processes and their mutual interactions.
Recently there has been impressive progress in the understanding of various types of relations between disease phenotypes 
on the basis of common underlying molecular processes \cite{Brev}.
For example, two diseases can be related if there are genes that are associated to both of them \cite{Goh,Feldman,Rzh,Park09}.
It was shown that genes associated with the same disorder encode proteins that have a strong tendency to interact with each other \cite{Goh}.
Alternatively, one can think of two diseases being linked and related if their metabolic reactions within a cell share common enzymes \cite{Lee}. 
Networks of protein-protein interactions \cite{barabsiN,barabsiN01} have also been studied in the context of disease interactions \cite{Ideker,BaCell}. 
Such interaction networks for seemingly unrelated gene products were shown to be involved in a group of different 
diseases that share clinical and pathological phenotypes \cite{BaCell}. 
The large number of relations between myriads of cellular components implies that diseases are not a clear clearcut concept, but 
merely act as some sort of a ''discretization'' in a vast and complicated phenotype space \cite{Brev}.
The structure of this space is typically studied through phenotypic comorbidity relations between human diseases.
A comorbidity relation means that two diseases occur more frequently within a patient than what would be expected from the 
frequency of the individual diseases alone. This means that the joint probability for suffering two diseases $i$ and $j$ together 
is larger than the product of the probabilities of the individual diseases (incidence rates), $P(i,j)>P(i)P(j)$. 
A phenotypic human disease network (PHDN) consists of nodes representing the diseases and links that indicate comorbidity relations \cite{Hid}.
PHDNs have opened a novel way for doing medicine on a population wide scale. For example
it has been shown that there exist pronounced racial differences in the PHDNs of black and white males \cite{Hid}, 
and that PHDNs may be used to predict future sites of cancer metastasis \cite{cancer}.
Note that only a limited number of comorbidities can be explained by common genes, proteins, or metabolites \cite{Rzh, Ch}.
These differences with respect to the clinical reality of diseases does not only reflect our limited knowledge of cellular processes, 
it also underscores the role of environmental and epigenetic factors in disease progressions.

The age-dependence of PHDNs is hitherto  unknown. Up to now studies on US Medicare data have uncovered 
comorbidity relations in patients aged 65 and older \cite{Hid,Ch}. In this work we use a complete medical claims data 
set that contains information on all of the 8.3 million Austrians that received medical treatments in the years 2006 and 2007.
The data set has been studied before to show a strong relationship between hunger in early life and the development of 
metabolic diseases in later life \cite{Diab, Gerontology}. Further, the nation-wide age- and gender dependence of diabetic 
complications was studied on the same data set \cite{leadleg}.

For the first time here we obtain a specific age-dependent PHDNs of all statistically significant comorbidity relations that pose a 
substantial risk to male or female patients. We do so by proposing a new statistical method that leads to an age- and gender specific
disease-disease network. This network is obtained by a combination of the layers of a disease-disease multiplex network that consists
of two layers that encode different phenomenological statistical measures for disease-disease relations.
In the first layer links quantify the statistical significance of a comorbidity relation between two diseases through their 
correlation coefficient for binary data. The technical challenge here is that the prevalences of individual diseases can 
vary over several orders of magnitude, from affecting a few dozens to ten thousands of patients in the database \cite{Hid}.  
This variability leads to biases in the correlations \cite{Olivier2013}.
In particular correlations between highly frequent and  rare diseases tend to be underestimated \cite{Hid, Olivier2013}.
We therefore employ a multi-scale correction which accounts for this bias \cite{Serrano}. Links between nodes $i$ and $j$ in 
the second layer represent the risk of obtaining disease $i$, given that the patient already suffers from disease $j$.
The first layer encodes information whether there exists a significant relation between two diseases,
the second layer quantifies the risk that a disease relation poses to the patient.

We quantify the topological network properties of the so obtained PHDNs and show that they undergo 
massive structural and gender-specific changes across lifetime. We show that this analysis allows us to understand the 
progression of the health state of a population on a new level. Instead of describing progressions of many individual diseases, 
it becomes apparent that various stages of life are characterized by a unique combination of tightly inter-related disease clusters.
With changing patient age these clusters of diseases emerge, vanish, merge, or form local hubs.
To a certain degree the concept of individual diseases becomes meaningless; what determines  disease risks and 
the health state of a population more effectively is the strongly age-dependent mesoscopic organization of the PHDN 
in disease clusters.

Finally, we develop a simple network diffusion model for the population-wide dynamics of disease progressions. 
The model is based on the empirical age- and gender-specific comorbidity relations recorded in the PHDN.
In particular we show that using (i) the prevalence of all diseases within a specific age group, and (ii) the age-dependent 
network structure of diseases for a given age, we can explain more than 85\% of the variance of the appearances of new  
diseases within the next eight years in the total population.
These results might provide important information for estimating the future burden of diseases in an aging society.
The fraction of the EU-population, aged 65 or older, will almost double by the year 2060 to 29.6\% of the population,
from 17.5\% in 2011. The average age of the EU-27 citizen is estimated to be 47.6 years, compared to  41.2 years in 2011 \cite{EUROPOP}.
However, quality of life is not determined by mere life expectancy {\em per se}, but by the number of years that are spent in healthy conditions.
Healthy-life years differ greatly between men and women and across member states of the EU.
Whereas life expectancy for women is 6.4y higher than for men, the number of healthy-life years for women is only 1.2y 
higher \cite{EHLE}.
The aging of population has severe implications for economic growth.
Estimates suggest that by 2030 the EU-27 will experience a 14\% decrease in workforce and a 7\% decrease in 
consumer population due to aging \cite{Hewitt02}.
It is therefore one of the big societal challenges to understand and anticipate 
to which extent the aging of the population affects the future health state of a population.

\section{\label{sec:dandm}Data and Methods}
\subsection{Data}
We use a database of the Main Association of Austrian Social Security Institutions that contains pseudonymized 
claims data of all persons receiving out- and inpatient care in Austria between January 1st, 2006 and December 31st, 2007. 
The data provides a comprehensive, nation-wide collection of the medical condition of the vast majority of Austria's population 
of 8.3 million people. Information on diagnosis is available for all persons receiving in-patient care, so-called inpatients.
The total sample of inpatients consists of 1,862,258 patients (1,064,952 females and 797,306 males) and includes their year of birth, 
gender, date of death, their drug prescriptions, and main- and side-diagnoses. 
Patients are grouped according to their age. The age groups, labeled by $a$, contain all patients whose age is between $a$ and $a+8$ years.
Diagnosis are provided in the 10th revision of the International Statistical Classification of Diseases and Related Health Problems 
(ICD 10) \cite{icd}, a medical classification list by the World Health Organization (WHO).
Not all ICD codes represent disorders, they may also indicate general examinations, injuries, pregnancies, or collections of symptoms  
that could be indicative of a large number of diseases. 
We exclude these categories and work with the $D = 1,055$ diagnoses on the three-digit 
ICD levels in chapters $A$-$N$.
We will use the words disease, disorder and diagnosis interchangeably whenever referring to a specific ICD entry.
It was found that such data can generally be considered to be complete 
with only a small amount of gross diagnostic miscoding \cite{Hen07}. Prognostic models for hospitalizations built on comorbidity 
scores derived from medical claims data perform similarly well as survey-derived data \cite{Zhang}.


 
\subsection{Derivation of the disease network from a statistical multiplex network}

We use the following notation. In our data we have $N^{m/f}(a)$ men/women in age group $a$.
Let us consider index pairs, where  index $i$ indicates patients suffering disease $i$, and $\neg i$ indicates patients not having the diagnosis $i$.
Each of the $N^{m/f}(a)$ patients belongs two one of four index pairs, $\{(i,j), (\neg i,j), (i,\neg j), (\neg i, \neg j) \}$, 
meaning that the patient has diseases $i$ and $j$,  has $j$ but not $i$, has $i$ but not $j$, or has neither $i$ nor $j$, respectively.
The numbers $N_{i,j}^{m/f}(a), N_{i,\neg j}^{m/f}(a),N_{\neg i, j}^{m/f}(a),N_{\neg i,\neg j}^{m/f}(a)$ denote the number of male/female 
patients in age group $a$ in these respective groups.
The {\it prevalence} for each disease $i$ is defined as $ p^{m/f}_i(a)=\frac{N^{m/f}_i(a)}{N^{m/f}(a)}$.
In the following we sometimes suppress the indices for age and gender for clarity.

\subsubsection{Layer 1: Statistical significance of comorbidity relations.}
The first layer of the PHDN multiplex provides the statistical significance of relations between any two diseases $i$ and $j$.
With the above notation the contingency coefficient $\phi_{ij}$  for a specific age group is defined by
\begin{equation}\label{phi}
\phi_{ij}=\frac{N_{i,j} N_{\neg i,\neg j} - N_{i,\neg j} N_{\neg i,j}}{\sqrt{N_i \cdot N_{\neg i} \cdot N_j \cdot N_{\neg j}}} \quad.
\end{equation}
A value of $\phi_{ij}(a)$ is computed for the patient numbers in each age group $a$.
These values are directly related to the $\chi$-square-coefficient $(\chi_{ij})^2 (a)= N(a) (\phi_{ij(a)})^2$, where $N(a)$ is the number of patients in the considered age group.
A chi-square test can be used to reject the null hypothesis for a given confidence level that the occurrence of disease $i$ is independent of the occurrence of disease $j$ in the same patient. 
$\phi = (\phi_{ij})$ is a symmetric matrix containing the test statistics for diseases $i$ and $j$.
We discard negative entries by setting $\phi_{ij} \to 0$, whenever $\phi_{ij} < 0$.
For later use we define an unweighted adjacency matrix $J$ derived from $\phi$, by defining $J_{ij}= 1$ if $\phi_{ij} > 0$ , 
and $J_{ij} = 0$ otherwise.

There are known biases in the test statistic $\phi_{ij}$ that depend on the prevalences $p^{m/f}_i(a)$ and $p^{m/f}_j(a)$ of the diseases $i$ and $j$.
While the contingency coefficient is a reliable measure for correlations whenever diseases $i$ and $j$ have similar prevalences $p^{m/f}_i(a)$ and $p^{m/f}_j(a)$,
$\phi$ underestimates correlations between very rare and highly frequent diseases \cite{Hid, Olivier2013}.
It is therefore unclear if low values for $\phi_{ij}$ for two diseases $i$ and $j$ imply weak correlations or large differences in their respective prevalences.
Assume that disease $i$ has a very low prevalence $p^{m/f}_i(a)$  with respect to the prevalence values of the majority of other diseases.
Consequently each of its $\phi$-values can be assumed to underestimate the true correlations, with respect to similar 
$\phi$-values obtained from more frequent diseases. 
A simple way to correct these biases is to compare values of $\phi_{ij}$ for disease $i$ with the typical correlation strength for disease $i$.
To this end consider the matrix 
\begin{equation}\label{chi}
\widetilde{\phi_{ij}}=\frac{\phi_{ij}}{\sum_{j=1}^D J_{ij}} \quad.
\end{equation}
While $\phi$ is an undirected network, $\widetilde{\phi}$ is typically directed.
To obtain a unique and bias corrected measure for the strength of the correlation between diseases $i$ and $j$ an undirected version $\phi^*$ of the network $\widetilde{\phi_{ij}}$ is finally obtained by 
$\phi^*_{ij} = \max  \left\{ \widetilde{\phi_{ij}},\widetilde{\phi_{ji}} \right\}$.

\subsubsection{Layer 2: Conditional disease risk.}
The second layer of the PHDN multiplex is given by the risk for a patient of given age group and gender to have disease 
$i$, given that she has disease $j$.
The conditional probability $P(i|j)$ for a patient who suffers disease $j$ to also suffer disease $i$ is given by $P(i|j) = \frac{N_{i,j}}{N_j}$.
A thresholded, undirected, and unweighted version $P^*$ of $P(i|j)$ is 
\begin{equation}\label{eq:p}
  P^{*}_{ij}= \left\{
    \begin{array}{ccl}
      1 &  & \max \left\{ P_{ij}, P_{ji}\right\}>p_c \\
      
      0 &  & {\rm otherwise} 
    \end{array}
  \right. \quad,
\end{equation}
for a given threshold $p_c>0$.

\subsubsection{Overlap of the multiplex network layers.}
The overlap (intersection) of the two multiplex network layers, $\phi*$ and $P^*$, the network $O$, is given by
\begin{equation}\label{eq:o}
  O_{i,j}= \left\{
    \begin{array}{cl}
      1 & \mathrm{if} \ \phi_{ij}^*>\phi_c \,\, and\,\,  P^*_{ij}>p_c  \\
      
      0 &  \mathrm{otherwise} 
    \end{array}
  \right. \quad.
\end{equation}
The network $O_{ij}$ links two diseases $i$ and $j$, if the link weights in both network 
layers exceed given thresholds, namely $\phi_c$ for the statistical significance layer $\Phi^*$,
and $p_c$ for the conditional risk layer $P^*$. 
Note that up to first order of $\frac{1}{\sum_{j=1}^D J_{ij}}$ the thresholding of the statistical significance layer $\phi^*$ by the value $\phi_c$, as done in eq. (\ref{eq:o}), yields the same results as the application of the multiscale filter proposed in \cite{Serrano}.
From now on we refer to  $O_{ij}$ simply as the ''disease network''. 
It is a variant of a PHDN. 
In contrast to the PHDN studied in \cite{Hid}, the disease network $O_{ij}$ is built around a combination of measures (conditional probabilities and contingency coefficients) that have been corrected
for biases that result from the comparison of very rare and frequent diseases.

\section{Results}

\subsection{Progression of diseases with aging.}

\begin{figure}[ht]
 \centerline{\epsfig{file=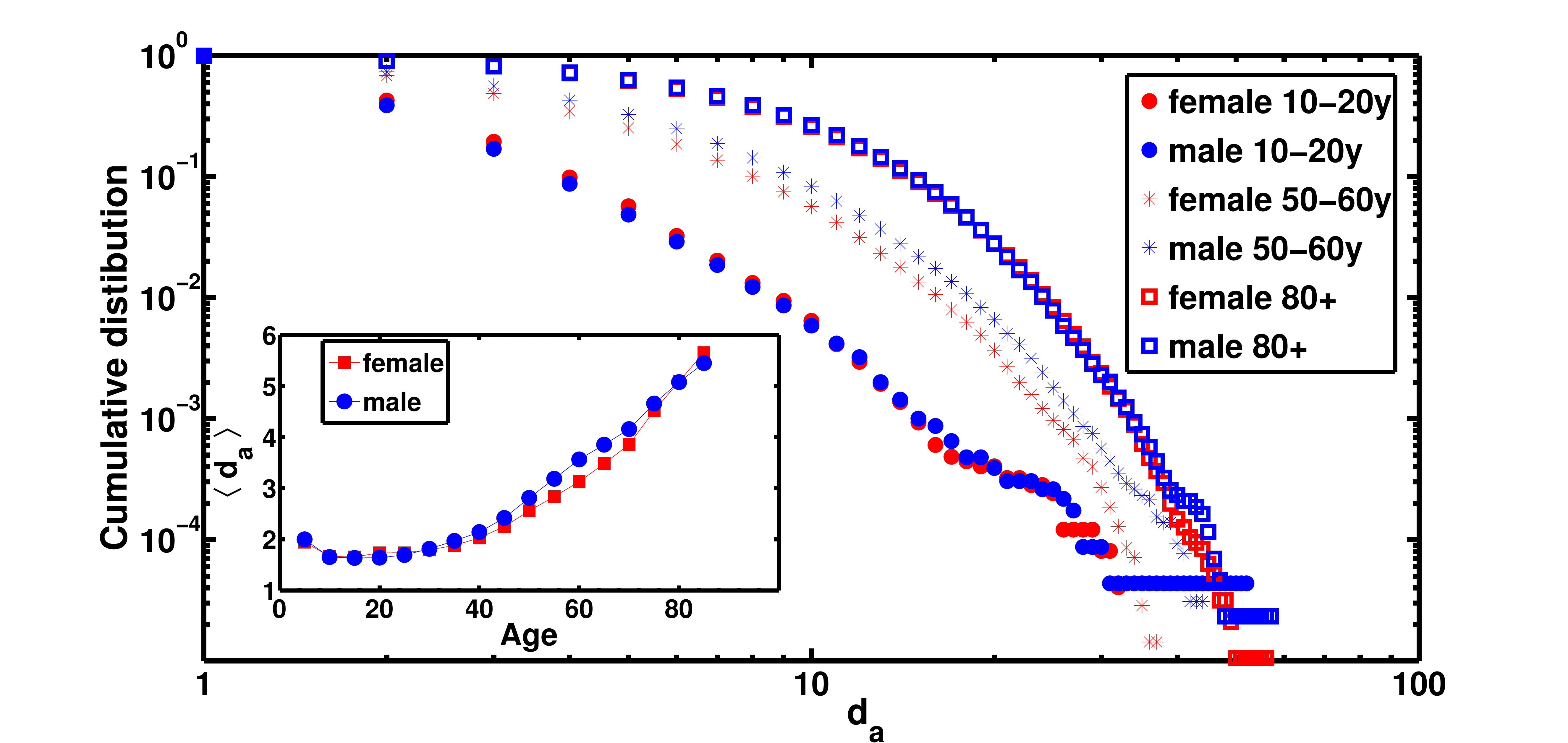,width=.8\columnwidth}}
    \caption{Cumulative distributions of the number of diagnosis for a typical inpatient for 
    three age groups $a=$10-20$y$, 50-60$y$, and $a>80y$ are shown. In the first age group the disease distributions 
    follow an approximate power-law. In older ages the distributions look more exponential. The inset shows the average 
    number of diagnoses $\langle d_a \rangle$ for inpatients for women (red  squares) and men (blue circles).}
   \label{diag}
\end{figure}

Let $d_a$ be the number of diseases diagnosed in patients with a hospital stay of at least one day in a given age group during the two years  2006 and 2007.
Figure \ref{diag} shows the cumulative distribution functions (cdc) for the number of diseases $d_a$ for three different age groups, 
$a=$10-20$y$, $a=$50-60$y$, and $a>80y$, for males and females.
The cdf for all age groups are presented in SI figures 1 and 2.
In the first age group the distribution suggests an approximate  power-law behavior up to about 20 diseases.
With increasing age the disease distributions takes on a more exponential character.
The average number of diseases per patient in each age group, $\langle d_a \rangle$, is shown in the inset of figure \ref{diag}.
Children in the first age group have a higher average number of diagnoses than teenagers.
After adolescence there is a continuous increase in the number of diseases per patient in the population, 
from two (at the age of around 20) to more than five for patients older than 80.
Men between 30-70 have generally more diseases then women, which reverses at older ages.
The number of diagnosis for females around 20 is higher than those for males.

\subsection{Network properties of the  disease network}

\begin{figure}[ht]
 \centerline{\epsfig{file=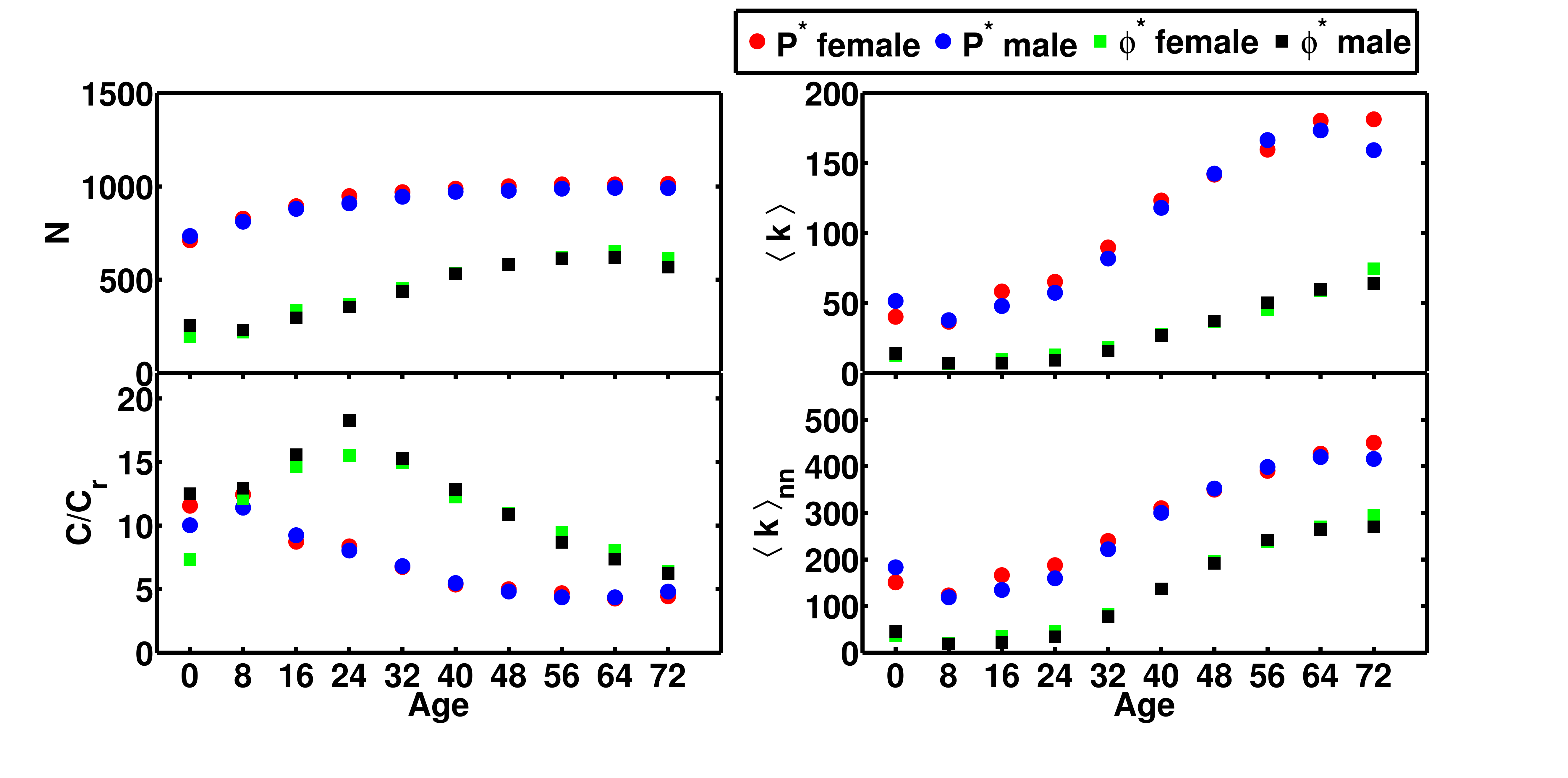,width=1.\columnwidth}}
    \caption{Basic network properties for the two levels $P^*$ and $\phi^*$ of the multiplex are compared across age. 
    Results for the conditional risk layer are shown in circles for females (red) and males (blue). For the statistical significance 
    layer $\phi^*$ results are shown for females (green) and males (black) in squares. (A) The number of nodes $N$ with at 
    least one link  increases in both network layers from childhood into adulthood, and levels 
    off at higher ages. (B) Average degrees  $\langle k \rangle$ increase over age. (C) Values of the clustering coefficients $C$ 
    divided by $C_r$ (clustering coefficient for corresponding random graph) show a pronounced peak around age 25 for the  
    $\phi^*$ network, and a consistent decrease in clustering for $P^*$. (D) Average degrees of nearest neighbors 
    $\langle k \rangle_{nn}$ increase with age. 
}
   \label{rys0}
\end{figure}

The large increase in diagnoses per patient over age is related to substantial topological changes of the $\phi^*$ and $P^*$ networks.
Network properties for the two layers are shown in figure \ref{rys0}.
The number of nodes $N$, figure \ref{rys0}(A), the average degrees $\langle k \rangle$, (B), and the average nearest neighbour 
degrees $\langle k \rangle_{nn}$, (D), increase with age in both layers.
For patients above 65 the degrees are larger for women than for men.
The clustering coefficients $C$ for each network are given in units of the clustering as expected 
from a random graph, $C_r$, in figure \ref{rys0}(C). 
$C/C_r$ is larger for $\phi$ than for the conditional risk layer for all age groups except for the youngest children. 
This ratio shows a clear maximum around age 25 in the $\phi^*$ layer, but decreases monotonically in the $P^*$ layer.

\begin{figure}[ht]
 \centerline{\epsfig{file=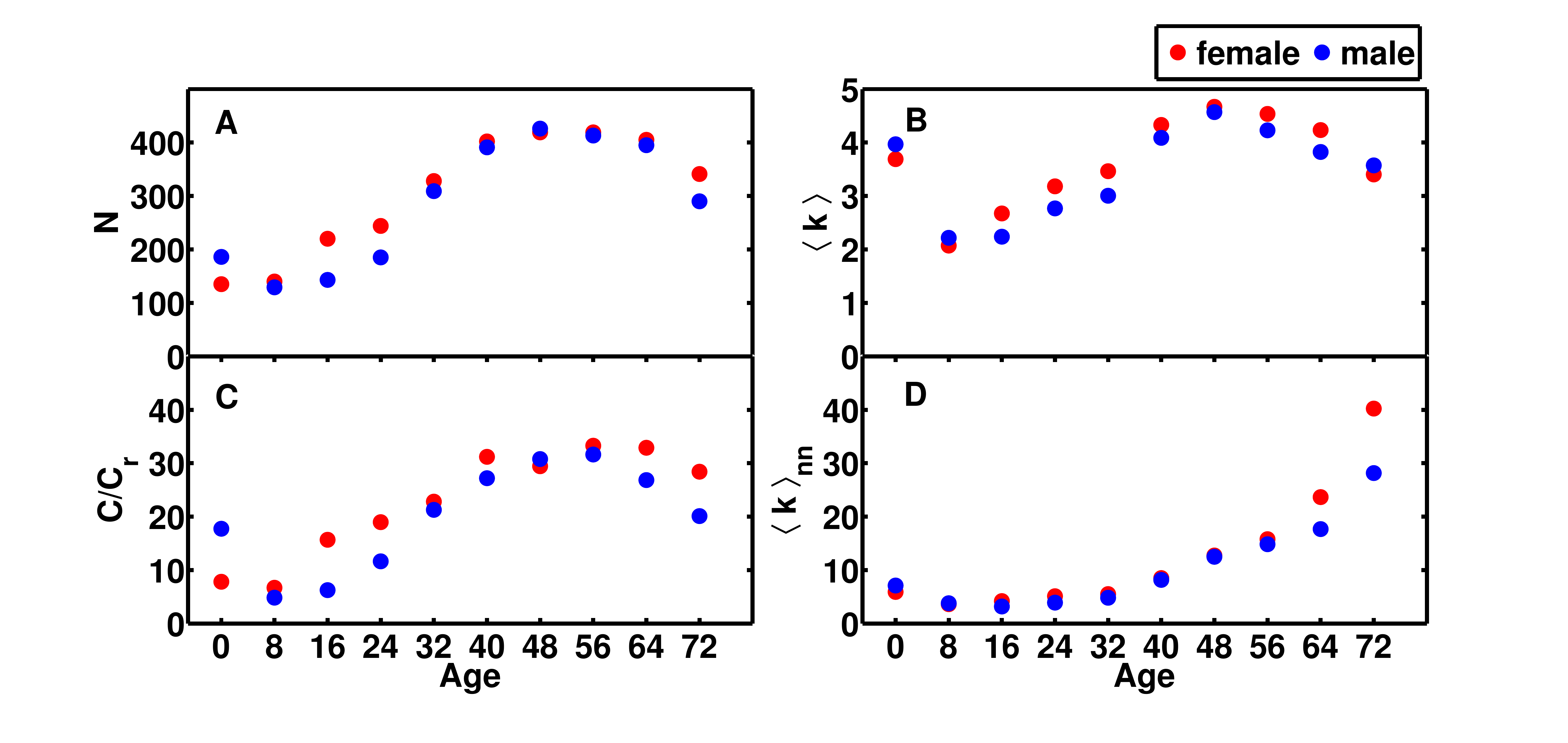,width=1.\columnwidth}}
    \caption{Network properties for the disease network $O$. (A)  number of nodes $N$, (B)  
    average degree $\langle  k \rangle$, (C)  clustering coefficients in units of $C_r$, and (D)  average nearest neighbour degrees $\langle k \rangle_{nn}$. 
   $O$ has the highest density for males and females aged $48-56$. Children are characterized by a higher density and higher 
   clustering than teenagers. There is a subsequent increase in $N$, $\langle k \rangle$ and $C/C_r$ with age up to a maximum 
   value at around age 50. The average degree of the nearest neighbors $\langle k \rangle_{nn}$ increases also for older ages.}
   \label{over}
\end{figure}

Network properties for the disease network $O$ for the threshold values $\phi_c=2$ and $p_c=0.01$ 
are shown in figure \ref{over}. The numbers of nodes $N$, figure \ref{over}(A), average degrees $\langle k \rangle$, (B), and the ratios 
of clustering coefficients $C/C_r$, (C), show a similar behavior with increasing age.
They decrease from the first to second age group, and increase until about age 50.
For higher ages the increase levels off and the network measures decrease again.
The average nearest neighbour degrees do not level off at high, but continue to increase, see figure \ref{over}(D).
The disease network is most dense for both males and females for the age range 48-56. 
We tested that the changes in network properties across age do not depend strongly on the choice of the thresholds $\phi_c$ and $p_c$.
We confirmed that these relations practically do not change when each threshold is altered by a factor of two.

Visualizations of the disease networks $O$ across lifetime are presented in figure \ref{viz} for males and females. 
Massive structural reorganisation  in the disease networks are clearly visible across age .
Nodes represent diseases $i$, their size is proportional to the disease prevalence, $N_i^{m/f}(a)$.
The  diseases type, as represented by the first letters from the ICD 10 code, is indicated by node colors.
Link colors are identical to the node colors if both diseases share the same type, otherwise it is a mix of the two colors. 
Disease clusters belonging to the same ICD type (first letter of ICD code) are highlighted by colored 
patches in figure \ref{viz}.

\subsection{Evolution of the human disease network across lifetime}

We find  three distinct phases in the evolution of the diseases networks across lifetime. 

\begin{figure}
\begin{center}
	\includegraphics[width=.85\columnwidth]{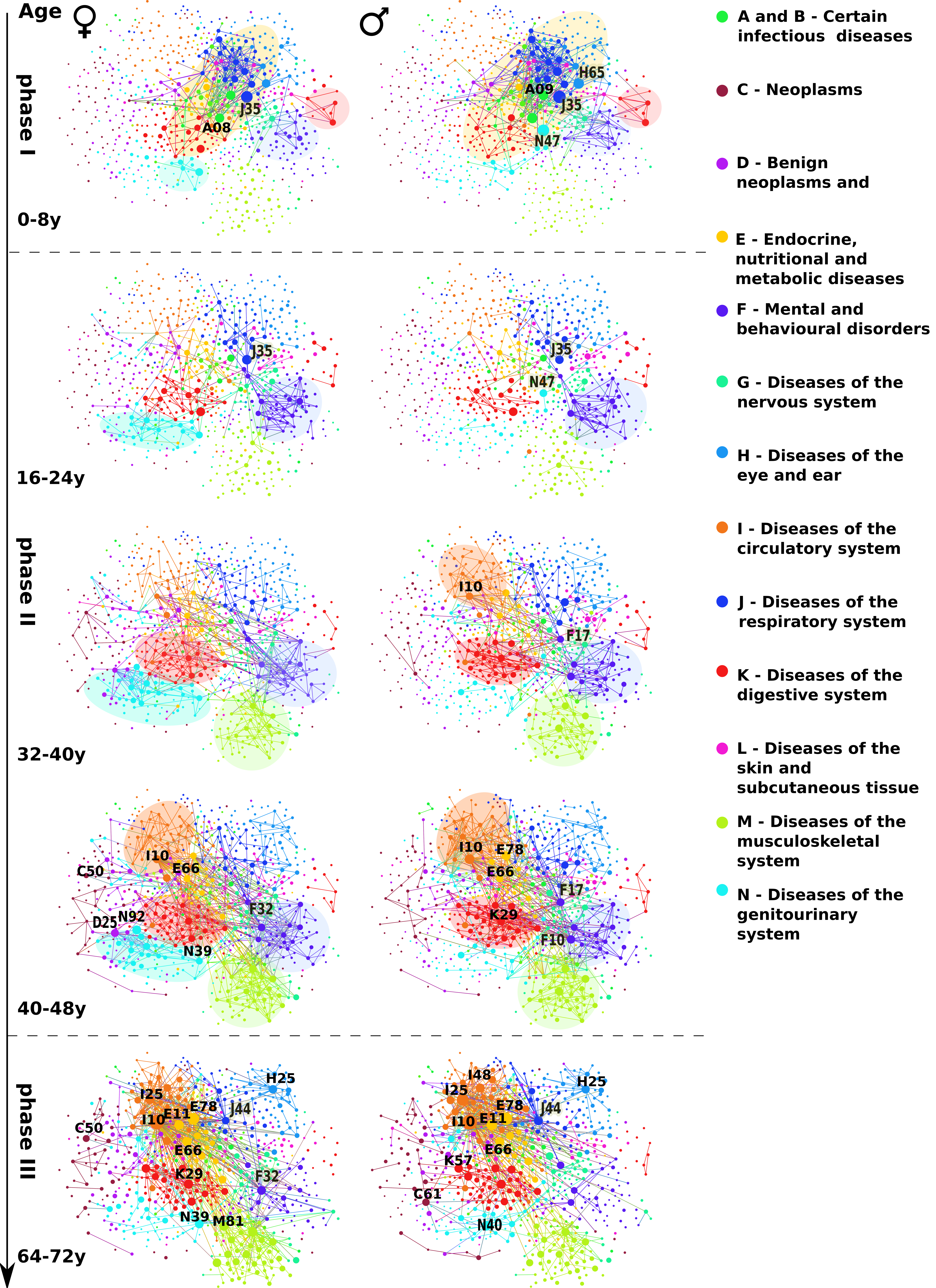}
\pagebreak	
	\end{center}
    \caption{Evolution of the disease network for females (left)  and  males (right). Each network corresponds 
    to a snapshot at a given age, with age increasing from top to bottom. Node sizes represent disease 
    prevalences, the colors indicate the main classifications of diseases according to the first digit of the ICD 10 codes. 
    The evolution of the disease network proceeds in roughly three phases: {\em Phase I,} children diseases, 
    characterized by one large cluster and several small ones. {\em Phase II} adult groups, in which many different 
    and clearly distinguishable clusters appear. {\em Phase III}, the network for elderly patients, where the network becomes 
    increasingly dominated by hubs. Colored patches serve as guides to the eye for clusters of diseases belonging to 
    the same type.}
   \label{viz}
\end{figure}

{\bf Phase I.} The disease network for children, phase I (age 0-16), shows one cluster containing diseases of  
many different types, such as diseases of the respiratory system (letter J in the ICD code), 
infectious diseases (A), diseases of the eye and ear (H), and endocrine, nutritional and metabolic diseases (E). 
Highly prevalent are viral and other specified intestinal infections (A08), diarrhoea and gastroenteritis 
of presumed infectious origin (A09), chronic diseases of tonsils and adenoids (J35), and non-suppurative otitis medi (H65).
These diseases are characterized by a large degree and tend to be connected to each other.
There is a small cluster formed by diseases of the nervous system (G), with a local hub being epilepsy 
(G40), and a cluster for mental and behavioral disorders (F). 
For male children there is a cluster containing disorders of psychological development that is not visible for females. 
For females in phase I there are small clusters containing diseases of the digestive system (K), and of the genitourinary system (N), 
whereas for males phimosis and paraphimosis (N47) belong to the main cluster. 
The network for young males is in general more dense than for females in phase I, 
but the formation of disease clusters and the prevalences are similar.
The network for patients aged 8-16 is less dense than the network of the former age group but the same clusters are still visible (see SI figure 7). 
There are obvious changes, however, for the next age groups 16-24, (see figure \ref{viz} second row), where the cluster for 
mental diseases (F) dominates and contains more different nodes than in the previous networks.
This can be seen as the onset of the structural phase II of the disease networks.

{\bf Phase II.} In phase II (age 16-64), a cluster of mental and behavioral disorders (F) appears more clearly and 
contains addiction to nicotine, alcohol, and other substance abuses. It has a higher density for women than for men. 
For both genders in the age group 16-24 the prevalences of diseases are smaller than in phase I, 
but infections (A08 and A09), tonsil problems (J35) and disorders of the prepuce for men (N47) are still highly prevalent.
Children and young patients suffer similarly from obesity (E66), which becomes increasingly prevalent starting with patients aged 16-24y. 
Hypertension (I10) becomes also visible at this age, but it is not yet a dominant disease.

At ages 32-40 the cluster of diseases of the digestive system (K) emerges for men (see figure \ref{viz} third row). 
Hemorrhoids (I84) are found at the periphery of this cluster with links exclusively to digestive diseases, 
there are no links from hemorrhoids to any other circulatory disease. 
The cluster of diseases of the genitourinary system (N) exists still only for women and grows with age.
It is connected to benign neoplasms and blood diseases (D).
For patients over 40 leiomyoma of uterus (D25) are very common.
This disease has many links to both genitourinary disorders (N) and to malignant neoplasms (C), 
especially related to female genital organs. 
The most prevalent neoplasms are breast cancer (C50) for women and prostate cancer (C61) for men. 
Neoplasms are located at the periphery of the network and show many inter-cluster connections, 
such as links to secondary neoplasms. 

The cluster of circulatory diseases (I) appears for men older than 32 and women older than 40. It 
densifies for the older age groups. This cluster is also much denser for men than for women, see figure \ref{viz}, fourth row. 
Another gender difference for this age group is the absence of genitourinary diseases (N) for men, 
while for women these diseases become increasingly prevalent until ages of 40.
The most common circulatory disease is hypertension (I10).
Already at ages 40-48 it has a large number of links inside and outside the cluster of circulatory diseases.

The origin of the cluster related to diseases of the musculoskeletal system (M) is visible for both 
genders at ages 24-32 (see SI figure 9); for older groups this cluster becomes much faster denser 
for females than for males. This may be related to osteoporosis (M81), which is a hub only for women 
aged over 48 and serves as a gateway to many musculoskeletal diseases. 
An important structural change begins at age 64 where some clusters start to shrink, such as mental 
disorders (F) and genitourinary diseases (N) for women. 
This marks the beginning of phase III (age 64-80) of the evolution of the disease network.

{\bf Phase III.} A contraction of the network towards its center can be observed in figure \ref{viz}, bottom row,  
where it becomes increasingly dominated by circulatory (I) and metabolic (E) diseases.
Global hubs appear, in particular hypertension (I10), depression (F32), ischemic heart diseases (I25), 
chronic obstructive pulmonary disease (COPD, J44), or the metabolic syndrome including disorders of 
lipoprotein metabolism (E78), type 2 diabetes (E11), and obesity (E66).
There are also some local hubs, for example cataracts (H25). 
Clearly discernible clusters disappear in phase III, with the exception of musculoskeletal disorders (M) for females.
Neoplasms (C) become increasingly interconnected.
The effect of the network contraction progresses in the oldest groups 72-80, 
where it is even more difficult to distinguish specific clusters (see SI figure 15).

The three phases are also apparent in the evolution of the network properties of the disease network, see figure \ref{over}.
In phase I (childhood) we find large clustering and degrees at a comparably small network size $N$.
This suggests a tightly interconnected cluster of diseases spanning only a small part of the entire possible network.
Phase II presents  itself as a steady increase in clustering and network density.
This trend levels off once substantial parts of the network show high levels of disease prevalences.
In phase III clustering and density decrease, whereas the average nearest neighbour degree increases.
This suggests the emergence of hubs which become connected to an increasing number of other diseases, 
whereas the local clusters that characterize phase II disappear.
On a more quantitative level we fitted the degree distributions with $q$-exponentials \cite{thurnertsallis,kyriakopolosthurnertsallis}, and report the 
values of $q$ for males and females across age groups in figures 2,3 and 4 in the SI. In general $q$ increases over lifetime, however there is a 
prominent peak at ages 16 and 24, indicating again a massive change of regime in network topology.

\section{Network diffusion model (NDM) of disease progression on the population level.}

It has been suggested that the spread of diseases may be related to a diffusion process on 
disease networks \cite{Hid}. Along these lines we propose a simple model that assumes that 
individuals can acquire new diseases that are comorbidities of already existing ones.
This is of course a crude simplification of the actual processes that lead to disease incidences.
  
Assume that the prevalence of disease $i$ in males/females of age group $a$ is $p_i^{m/f}(a) = \frac{N_i^{m/f}(a)}{N^{m/f}(a)}$.
Let the expected prevalence of disease $i$ at the next age group be denoted by $\hat p_i^{m/f}(a+1)$.
We then make the simple network diffusion Ansatz, 
\begin{equation}\label{delta}
 \hat p_i^{m/f}(a+1)=p_i^{m/f}(a)+(1-p_i^{m/f}(a))\sum_{j \neq i} \Delta P^{m/f}_{ij}(a)p_j^{m/f}(a) \quad, 
 \label{model2}
\end{equation}
where $\Delta P^{m/f}_{ij}(a)=P^{m/f}_{ij}(a+1)-P^{m/f}_{ij}(a)$ is the difference between the conditional 
disease risks for age groups $a+1$ and $a$. $\Delta P^{m/f}_{ij}(a)$ is the risk of obtaining disease $i$ 
given $j$ for a patient aging from age group $a$ to $a+1$. This number is multiplied with the probabilities 
to have disease and to {\em not} have disease $i$ to give the disease-network-effect on the disease prevalences.

\begin{figure}[ht]
 \centerline{\epsfig{file=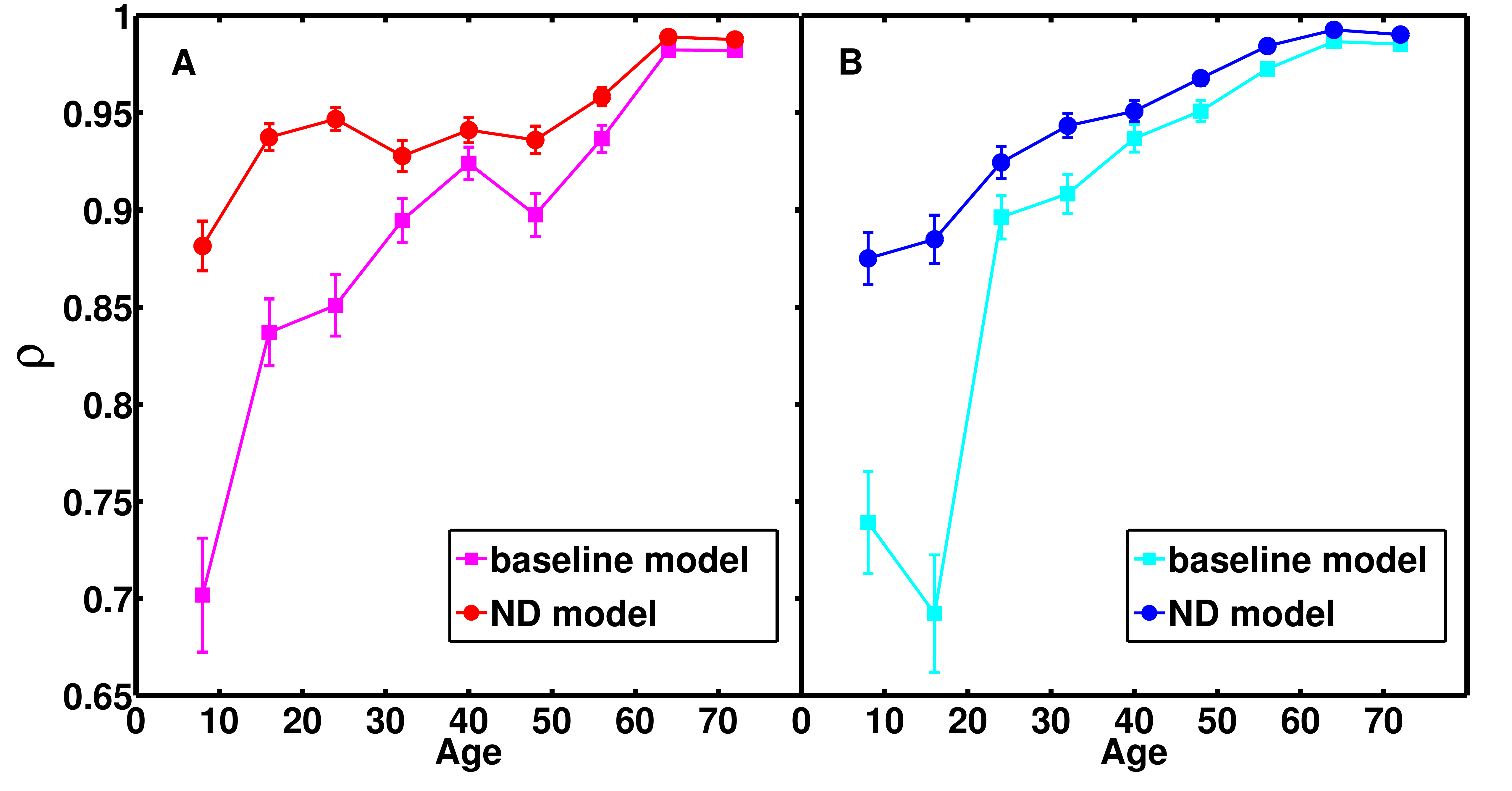,width=.8\columnwidth}}
    \caption{Correlation coefficients $\rho$ between actual prevalence data and  predictions from the network diffusion model (NDM)
    are shown for  females (A) and males (B). NDM results are compared to the baseline model without diffusion term.
    The disease-network-effect (difference between lines) consistently 
    increases correlations up to 20\% especially for children and adolescents. The importance of diffusion diminishes with higher ages. }
   \label{model}
\end{figure}

To estimate the quality of the model we compute the  correlation coefficient $\rho$ between the actually observed 
prevalences at the next timestep $p_i^{m/f}(a+1)$,  and the predicted result of equation \ref{model2}, 
$\rho= \mathrm{corr}\left( \hat p_i^{m/f}(a+1),p_i^{m/f}(a+1)\right)$. The value of $\rho$ quantifies how much of the 
variance of disease prevalences at a given age $a$  can be explained by the prevalences of diseases in the previous 
age group, superimposed by the risk from disease spreading from nearby nodes in the disease network.
The values for $\rho$ for the network diffusion model (NDM) are shown in figure \ref{model}(A) for women and (B) for men. 
The correlation coefficients for the NDM are higher than 90\% for ages above 50. 
To estimate the importance of comorbidities in the NDM, we compare its results with the correlation 
coefficient obtained for a variant of the model in equation (\ref{model2}), where 
we ignore the second term on the right hand side, i.e. we ignore effects from the disease network and set $\Delta P^{m/f}_{ij}=0$. 
This variant we call the ''baseline model''. Results are also collected in figure \ref{model}. 
These results suggest that a substantial fraction of disease incidents can be predicted on the basis of the 
diseases that exist at the previous age group alone. 
Clearly, the diffusion model consistently outperforms the baseline model for all ages and gender groups.
In the age range 40-80 the disease-network-effect increases the explanatory power compared to the baseline 
model only by a couple of percent. Gender differences can be observed. For women at age 48 the quality gap between 
the models is about 5\%, compared to 2-3\% for men. This may be related to physiological changes during the menopause. 
Important physiological changes also take place during puberty and adolescence. During these ages the baseline model 
is able to explain up to 70\% of the variance of disease prevalences, whereas the NDM performs substantially better 85-90\%. 

\section{\label{sec:discussion}Discussion}

The health state of patients is typically characterized by multiple comorbid conditions, especially among adults and the elderly.
We proposed a phenotypic human disease network that is derived from a multiplex network of two statistical 
measures quantifying relations between diseases. We showed that the disease network undergoes dramatic structural 
transformations as a function of patient age and gender. In particular one can distinguish three phases in the evolution of the disease network.
The first phase describes the comorbidity relations in children and consists of a single, strongly inter-connected cluster. 
The second phase for adolescents and adults is characterized by the appearance of subsequent clusters related to specific classes of diseases.
For instance, clusters of mental disorders appear in both genders in younger adults, whereas genitourinary disorders appear only for women. 
Clusters for circulatory, musculoskeletal, and digestive diseases appear in subsequent life years.
In the last phase for elderly  people above 65, disease clusters condense and a several highly connected hubs appear and dominate the network. 
These hubs are  hypertension, chronic ischemic heart diseases, and COPD .
The results presented here allow to understand developments in population health across life in a new way.
With the adoption of a network perspective, the focus shifts from the study of individual disease progression to transformations 
of the mesoscopic organization of the disease network. We showed that each stage of life can be characterized by its unique set 
of clusters of closely inter-related diseases. We developed a network diffusion model which shows that patients indeed develop 
new diseases in close proximity in the disease network to disorders they already have. A thorough topological understanding of 
disease-disease relations is therefore key in order to anticipate future developments in population health.
It is remarkable that the vast majority of new disease onsets can be explained by the conditional disease risks $P(i|j)$ alone.
Combined with models that describe demographic changes in the age structure of a population, the network diffusion model proposed in this work may provide an attractive starting point to predict future burdens of diseases within an aging society.


\section{Acknowledgments}
A.Ch. was supported by the EU FP7 project LASAGNE no. 318132, P.K. by EU FP7 project  MULTIPLEX  No. 317532



\end{document}